\documentclass[10pt]{article}
\newtheorem{theorem}{Theorem}[section]
\newtheorem{lemma}[theorem]{Lemma}
\newtheorem{remark}[theorem]{Remark}
\newtheorem{definition}[theorem]{Definition}
\newtheorem{corollary}[theorem]{Corollary}

\usepackage{amsmath}
\usepackage{amsfonts}
\usepackage{amstext}
\usepackage{amssymb}
\usepackage{setspace}

\begin{document}
\title{Asymptotic Formulas with Error Estimates for Call Pricing Functions and the Implied Volatility at Extreme Strikes}
\date{}
\author{Archil Gulisashvili}
\maketitle
\vspace{0.2in}
\bf Abstract \rm In this paper, we obtain asymptotic formulas with error estimates for the implied volatility associated with a European call pricing function. 
We show that these formulas imply Lee's moment formulas for the implied volatility and the tail-wing formulas due to Benaim and Friz. 
In addition, we analyze Pareto-type tails of stock price distributions in uncorrelated Hull-White, Stein-Stein, and Heston models and find asymptotic formulas with error estimates for 
call pricing functions in these models.
\\
\\
\bf Keywords\,\, \rm Call and put pricing functions $\cdot$ Implied volatility $\cdot$ Asymptotic formulas $\cdot$ Pareto-type distributions $\cdot$ Regularly varying functions
\hspace{1in} 
\normalsize

\footnotetext{A. Gulisashvili \\
Department of Mathematics, Ohio University, Athens, OH 45701, USA \\
e-mail: guli@math.ohiou.edu}
\section{Introduction}\label{S:in}
In this paper, we study the asymptotic behavior of the implied volatility $K\mapsto I(K)$ associated with a 
European type call pricing function $K\mapsto C(K)$. Here the symbol $K$ stands for the strike price, and it is assumed that the  
expiry $T$ is fixed. 
One of the main results obtained in the present paper is the following asymptotic formula, which is true for 
any call pricing function $C$:
\begin{equation}
I(K)=\frac{\sqrt{2}}{\sqrt{T}}\left[\sqrt{\log K+\log\frac{1}{C(K)}}-\sqrt{\log\frac{1}{C(K)}}\right] 
+O\left(\left(\log\frac{1}{C(K)}\right)^{-\frac{1}{2}}\log\log\frac{1}{C(K)}\right)
\label{E:ai1}
\end{equation}
as $K\rightarrow\infty$. A similar formula holds for $K$ near zero: 
\begin{equation}
I(K)=\frac{\sqrt{2}}{\sqrt{T}}\left[\sqrt{\log\frac{1}{P(K)}}-\sqrt{\log\frac{1}{P(K)}-\log\frac{1}{K}}\right]
+O\left(\left(\log\frac{K}{P(K)}\right)^{-\frac{1}{2}}\log\log\frac{K}{P(K)}\right)
\label{E:ai2}
\end{equation}
as $K\rightarrow 0$, where $K\mapsto P(K)$ is the put pricing function corresponding to $C$. In Sections \ref{S:lee} and \ref{S:cBF}, we will compare 
formulas (\ref{E:ai1}) and (\ref{E:ai2}) with known asymptotic formulas
for the implied volatility. For instance, it will be shown that Lee's moment formulas (see \cite{keyL2}) and the 
tail-wing formulas due to Benaim and Friz (see \cite{keyBF}) can be derived using 
(\ref{E:ai1}) and (\ref{E:ai2}). 

Let $X$ be a positive adapted stochastic process defined on a filtered probability space 
$\left(\Omega,{\cal F},{\cal F}_t,\mathbb{P}^{*}\right)$. The process $X$ models the random behavior of the stock price.
It is assumed that for every $t> 0$, $X_t$ is an unbounded random variable and that the process $X$ satisfies the following conditions:
$X_0=x_0$ $\mathbb{P}^{*}$-a.s. for some $x_0>0$ and $\mathbb{E}^{*}\left[X_t\right]<\infty$ for every $t\ge 0$.
In addition, we suppose that $\mathbb{P}^{*}$ is a risk-free measure. This means that the discounted stock price process $\left\{e^{-rt}X_t\right\}_{t\ge 0}$ is 
a $\left({\cal F}_t,\mathbb{P}^{*}\right)$-martingale. Here $r\ge 0$ denotes the interest rate. It follows that
\begin{equation}
x_0=e^{-rt}\mathbb{E}^{*}\left[X_t\right],\quad t\ge 0.
\label{E:mapr}
\end{equation}
Under these conditions, the pricing function for a European call option at time $t=0$ is defined by
\begin{equation}
C(T,K)=e^{-rT}\mathbb{E}^{*}\left[\left(X_T-K\right)^{+}\right]
\label{E:pfc}
\end{equation}
where $K\ge 0$ is the strike price, $T\ge 0$ is the expiry, and for every real number $u$, $u^{+}=\max\{u,0\}$. 
If $C$ is a call pricing function, then the corresponding put pricing function $P$ is defined by
\begin{equation}
P(T,K)=e^{-rT}\mathbb{E}^{*}\left[\left(K-X_T\right)^{+}\right].
\label{E:pfp}
\end{equation}
The functions $C$ and $P$ satisfy the put-call parity condition 
$C(T,K)=P(T,K)+x_0-e^{-rT}K$. 
This formula can be easily derived from (\ref{E:pfc}) and (\ref{E:pfp}).

When is a positive function of two variables a call pricing function at time $t=0$ with 
interest rate $r$ and initial condition $x_0$? The answer to this question formulated below is a combination of several known statements. However, we found only 
less general results in the literature.
For instance, Carmona and Nadtochiy (see \cite{keyCN}) describe conditions which are imposed on call pricing functions by the absence of arbitrage without giving details. 
The description in \cite{keyCN} essentially coincides with the necessity part of the assertion formulated below. We also refer the reader to Section 3 of the paper \cite{keyB} 
of Buehler where similar results are obtained. We will prove the characterization theorem in the appendix. The proof is included for the sake of completeness.
Kellerer's theorem that appears in the proof concerns marginal distributions of Markov 
martingales (see \cite{keyK}, see also \cite{keyHK,keyMY}). This theorem is often used in the papers devoted to the 
existence of option pricing models reproducing observed option prices (see \cite{keyB,keyCM,keyC} 
and the references therein, see also \cite{keyDH}, where the Sherman-Stein-Blackwell theorem is employed instead of Kellerer's theorem). 

The next assertion characterizes general call pricing functions. Let $C$ be a strictly positive function on $[0,\infty)^2$. 
Then $C$ is a call pricing function if and only if the following conditions hold:
\begin{enumerate}
\item For every $T\ge 0$, the function $K\rightarrow C(T,K)$ is convex.
\item For every $T\ge 0$, the second distributional derivative $\mu_T$ of the function $K\mapsto e^{rT}C(T,K)$ is a Borel probability measure such that
\begin{equation}
\int_0^{\infty}xd\mu_T(x)=x_0e^{rT}.
\label{E:vo}
\end{equation}
\item For every $K\ge 0$, the function $T\rightarrow C(T,e^{rT}K)$ is non-decreasing.
\item For every $K\ge 0$, $\displaystyle{C(0,K)=\left(x_0-K\right)^{+}}$.
\item For every $T\ge 0$, $\displaystyle{\lim_{K\rightarrow\infty}C(T,K)=0}$.
\end{enumerate}

A popular example of a call pricing function is the function $C_{BS}$ arising in the Black-Scholes model. In this model, the stock price process is a geometric Brownian motion, 
satisfying the stochastic differential equation $dX_t=rX_tdt+\sigma X_tdW_t^{*}$, where $r\ge 0$ is the interest rate, $\sigma> 0$ is the volatility of the stock, 
and $W^{*}$ is a standard Brownian motion under the risk-free measure $\mathbb{P}^{*}$. The process $X$ is given by
\begin{equation}
X_t=x_0\exp\left\{\left(r-\frac{\sigma^2}{2}\right)t+\sigma W^{*}_t\right\}
\label{E:fof}
\end{equation}
where $x_0> 0$ is the initial condition. Black and Scholes found an explicit formula for the pricing function $C_{BS}$. This formula is as follows:
\begin{equation}
C_{BS}\left(T,K,\sigma\right)=x_0N\left(d_1(K,\sigma)\right)-Ke^{-rT}N\left(d_2(K,\sigma)\right),
\label{E:BSF}
\end{equation}
where
$$
d_1(K,\sigma)=\frac{\log x_0-\log K+\left(r+\frac{1}{2}\sigma^2\right)T}
{\sigma\sqrt{T}},\quad d_2(K,\sigma)=d_1(K,\sigma)-\sigma\sqrt{T},
$$
and 
$$
N(z)=\frac{1}{\sqrt{2\pi}}\int_{-\infty}^z\exp\left\{-\frac{y^2}{2}\right\}dy.
$$
We refer the reader to \cite{keyLL} for more information on this celebrated result.

Let $C$ be a general call pricing function. The implied volatility 
$I=I(T,K)$, $(T,K)\in[0,\infty)^2$, associated with the pricing function $C$, is a function of two variables satisfying the following condition:
$$
C_{BS}(T,K,I(T,K))=C(T,K).
$$ 
It is well-known that for every pair $(T,K)\in[0,\infty)^2$, the number $I(T,K)$ for which the previous equality holds, exists and is unique. 
We refer the reader to \cite{keyF,keyFPS,keyG,keyHL} for additional information on the implied volatility. 
The asymptotic behavior of the implied volatility for extreme strikes was studied in \cite{keyBF,keyBF1,keyBFL,keyGS2,keyGS4,keyL2} (see also Sections 10.5 and 10.6 of \cite{keyHL}). 

In the present paper, various asymptotic relations between functions are exploited. 
\begin{definition}\label{D:ae} 
Let $\varphi_1$ and $\varphi_2$ be positive functions on $(a,\infty)$. We define several asymptotic relations by the following:
\begin{enumerate}
\item If there exist $\alpha_1> 0$, $\alpha_2> 0$, and $y_0> 0$ such that 
$\alpha_1\varphi_1(y)\le\varphi_2(y)\le \alpha_2\varphi_1(y)$ for all $y> y_0$,
then we use the notation $\varphi_1(y)\approx\varphi_1(y)$ as $y\rightarrow\infty$. 
\item If $\lim_{y\rightarrow\infty}\left[\varphi_2(y)\right]^{-1}\varphi_1(y)=1$,
then we write $\varphi_1(y)\sim\varphi_2(y)$ as $y\rightarrow\infty$.
\item Let $\rho$ be a positive function on $(0,\infty)$. We use
the notation
$\varphi_1(y)=\varphi_2(y)+O(\rho(y))$ as $y\rightarrow\infty$,
if there exist $\alpha> 0$ and $y_0> 0$ such that
$\left|\varphi_1(y)-\varphi_2(y)\right|\le\alpha\rho(y)$ for all $y> y_0$.
\end{enumerate}
\end{definition}
Similar definitions will be used in the case where $y\downarrow 0$.

We will next give a quick overview of the results obtained in the present paper.  
In Sections \ref{S:BS} and \ref{S:kdo}, we find various asymptotic formulas for the implied volatility associated with 
a general call pricing function. In Section \ref{S:lee}, we give a new proof of Lee's moment formulas 
for the implied volatility, while in Section \ref{S:cBF} we compare our asymptotic formulas with the tail-wing formulas due to
Benaim and Friz. We also obtain tail-wing formulas with error estimates under additional restrictions. In Section \ref{S:exam}, we talk about Pareto-type tails of stock price distributions in 
uncorrelated Hull-White, Stein-Stein, and Heston models. For these distributions, we compute the Pareto-type index and find explicit expressions for the corresponding slowly varying 
functions. In Section \ref{S:ssvm}, 
we obtain sharp asymptotic formulas for pricing functions in uncorrelated Hull-White, Stein-Stein, and Heston models. 
Finally, in the appendix we prove the characterization theorem for call pricing functions formulated in the introduction.
\section{Asymptotic behavior of the implied volatility as $K\rightarrow\infty$}\label{S:BS}
In this section, we find sharp asymptotic formulas for the implied volatility $K\mapsto I(K)$ associated with a general pricing function $C$. 
Recall that the following conditions hold for any call pricing function: $C(K)\rightarrow 0$ as $K\rightarrow\infty$ and $C(K)> 0$ for all $K> 0$.
\begin{theorem}\label{T:generu}
Suppose that $C$ is a call pricing function, and let $\psi$ be a positive function with 
$\displaystyle{\lim_{K\rightarrow\infty}\psi(K)=\infty}$. Then
\begin{align}
&I(K)=\frac{1}{\sqrt{T}}\left[\sqrt{2\log K+2\log\frac{1}{C(K)}-\log\log\frac{1}{C(K)}}
-\sqrt{2\log\frac{1}{C(K)}-\log\log\frac{1}{C(K)}}\right]
\nonumber \\
&\quad+O\left(\left(\log\frac{1}{C(K)}\right)^{-\frac{1}{2}}\psi(K)\right)
\label{E:repl1}
\end{align}
as $K\rightarrow\infty$. 
\end{theorem}

Theorem \ref{T:generu} and the mean value theorem imply the following assertion:
\begin{corollary}\label{C:geu}
For any call pricing function $C$,
\begin{equation}
I(K)=\frac{\sqrt{2}}{\sqrt{T}}\left[\sqrt{\log K+\log\frac{1}{C(K)}}
-\sqrt{\log\frac{1}{C(K)}}\right]
+O\left(\left(\log\frac{1}{C(K)}\right)^{-\frac{1}{2}}\log\log\frac{1}{C(K)}\right)
\label{E:repli}
\end{equation}
as $K\rightarrow\infty$. 
\end{corollary}
\it Proof of Theorem \ref{T:generu}. \rm 
The following lemma was established in \cite{keyGS2,keyGS4} under certain restrictions on the pricing function. 
The proof in the general case is similar. 
\begin{lemma}\label{L:gene2}
Let $C$ be a call pricing function and fix a positive continuous increasing function $\psi$ with 
$\displaystyle{\lim_{K\rightarrow\infty}\psi(K)=\infty}$. Suppose that $\phi$ 
is a positive Borel function such that $\displaystyle{\lim_{K\rightarrow\infty}\phi(K)=\infty}$ and 
$$
C(K)\approx\frac{\psi(K)}{\phi(K)}\exp\left\{-\frac{\phi(K)^2}{2}\right\}.
$$
Then the following asymptotic formula holds:
$$
I(K)=\frac{1}{\sqrt{T}}\left(\sqrt{2\log\frac{K}{x_0e^{rT}}+\phi(K)^2}-\phi(K)\right)+O\left(\frac{\psi(K)}{\phi(K)}\right)
$$
as $K\rightarrow\infty$.
\end{lemma}

With no loss of generality, we can assume that the function $\psi(K)$ tends to 
infinity slower than the function 
$\displaystyle{K\mapsto\log\log\frac{1}{C(K)}}$. Put 
$$
\phi(K)=\left[2\log\frac{1}{C(K)}-\log\log\frac{1}{C(K)}+2\log\psi(K)\right]^{\frac{1}{2}}.
$$
Then we have
$\displaystyle{\phi(K)\approx\sqrt{2\log\frac{1}{C(K)}}}$ as $K\rightarrow\infty$
and it follows that
$$
\psi(K)\exp\left\{-\frac{\phi(K)^2}{2}\right\}\phi(K)^{-1}\approx C(K),\quad K\rightarrow\infty.
$$
Therefore, Lemma \ref{L:gene2} gives
\begin{equation}
I(K)=\frac{1}{\sqrt{T}}\left(\sqrt{2\log\frac{K}{x_0e^{rT}}+\phi(K)^2}-\phi(K)\right)
+O\left(\left(\log\frac{1}{C(K)}\right)^{-\frac{1}{2}}\psi(K)\right)
\label{E:gener3}
\end{equation}
as $K\rightarrow\infty$. Now, it is not hard to see that (\ref{E:gener3}) and the mean value theorem imply (\ref{E:repl1}).

This completes the proof of Theorem \ref{T:generu}.

Our next goal is to replace the function $C$ in formula (\ref{E:repl1}) by another function $\widetilde{C}$. 
\begin{corollary}\label{C:gik1}
Let $C$ be a call pricing function and let $\psi$ be a positive function with 
$\displaystyle{\lim_{K\rightarrow\infty}\psi(K)=\infty}$. Suppose that $\widetilde{C}$ is a positive function for which
$\widetilde{C}(K)\approx C(K)$ as $K\rightarrow\infty$. Then
\begin{align}
&I(K)=\frac{1}{\sqrt{T}}\left[\sqrt{2\log K+2\log\frac{1}{\widetilde{C}(K)}-\log\log\frac{1}{\widetilde{C}(K)}}
-\sqrt{2\log\frac{1}{\widetilde{C}(K)}-\log\log\frac{1}{\widetilde{C}(K)}}\right]
\nonumber \\
&\quad+O\left(\left(\log\frac{1}{\widetilde{C}(K)}\right)^{-\frac{1}{2}}\psi(K)\right)
\label{E:gik1}
\end{align}
as $K\rightarrow\infty$. In addition,
\begin{equation}
I(K)=\frac{\sqrt{2}}{\sqrt{T}}\left[\sqrt{\log K+\log\frac{1}{\widetilde{C}(K)}}
-\sqrt{\log\frac{1}{\widetilde{C}(K)}}\right]+O\left(\left(\log\frac{1}{\widetilde{C}(K)}\right)^{-\frac{1}{2}}\log\log\frac{1}{\widetilde{C}(K)}\right)
\label{E:gikko}
\end{equation}
as $K\rightarrow\infty$.
\end{corollary}

Formula (\ref{E:gik1}) can be established exactly as (\ref{E:repl1}). Formula (\ref{E:gikko}) follows from (\ref{E:gik1}) and the mean value theorem.

We can also replace a call pricing function $C$ in (\ref{E:repl1}) by a function $\widetilde{C}$ under more general conditions. However, this may lead to a weaker error estimate.
Put
\begin{equation}
\tau(K)=\left|\log\frac{1}{C(K)}-\log\frac{1}{\widetilde{C}(K)}\right|.
\label{E:tau}
\end{equation}
Then the following theorem holds:
\begin{theorem}\label{T:gener}
Let $C$ be a call pricing function and $\psi$ be a positive function with
$\displaystyle{\lim_{K\rightarrow\infty}\psi(K)=\infty}$. Suppose $\widetilde{C}$ is a positive function satisfying the following condition:
There exist $K_1> 0$ and $c$ with $0< c< 1$ such that
\begin{equation}
\tau(K)<c\log\frac{1}{\widetilde{C}(K)}\quad\mbox{for all}\quad K> K_1,
\label{E:ree}
\end{equation}
where $\tau$ is defined by (\ref{E:tau}). Then
\begin{align}
&I(K)=\frac{1}{\sqrt{T}}\left[\sqrt{2\log K+2\log\frac{1}{\widetilde{C}(K)}-\log\log\frac{1}{\widetilde{C}(K)}}
-\sqrt{2\log\frac{1}{\widetilde{C}(K)}-\log\log\frac{1}{\widetilde{C}(K)}}\right]
\nonumber \\
&\quad+O\left(\left(\log\frac{1}{\widetilde{C}(K)}\right)^{-\frac{1}{2}}[\psi(K)+\tau(K)]\right)
\label{E:repl3}
\end{align}
as $K\rightarrow\infty$. 
\end{theorem}

\it Proof. \rm It is not hard to check that (\ref{E:ree}) implies the formula
$\displaystyle{\log\frac{1}{\widetilde{C}(K)}\approx\log\frac{1}{C(K)}}$ as $K\rightarrow\infty$. Now using (\ref{E:repl1}), 
(\ref{E:tau}), and the mean value theorem, we obtain (\ref{E:repl3}).

The next statement follows from the special case of Theorem \ref{T:gener} where $\displaystyle{\psi(K)=\log\log\frac{1}{\widetilde{C}(K)}}$ 
and from the mean value theorem.
\begin{corollary}\label{C:gik2}
Let $C$ be a call pricing function, and suppose $\widetilde{C}$ is a positive function satisfying the following condition:
There exist $\nu> 0$ and $K_0> 0$ such that
\begin{equation}
\left|\log\frac{1}{\widetilde{C}(K)}-\log\frac{1}{C(K)}\right|\le\nu\log\log\frac{1}{\widetilde{C}(K)},\quad K> K_0.
\label{E:sv}
\end{equation}
Then
\begin{equation}
I(K)=\frac{\sqrt{2}}{\sqrt{T}}\left[\sqrt{\log K+\log\frac{1}{\widetilde{C}(K)}}-\sqrt{\log\frac{1}{\widetilde{C}(K)}}\right]
+O\left(\left(\log\frac{1}{\widetilde{C}(K)}\right)^{-\frac{1}{2}}\log\log\frac{1}{\widetilde{C}(K)}\right)
\label{E:repl4}
\end{equation}
as $K\rightarrow\infty$. 
\end{corollary}

It is not hard to see that if $C(K)\approx\widetilde{C}(K)$ as $K\rightarrow\infty$, then 
$\displaystyle{\log\frac{1}{C(K)}\sim\log\frac{1}{\widetilde{C}(K)}}$
as $K\rightarrow\infty$. The previous formula also follows from (\ref{E:sv}).
\begin{corollary}\label{C:geg}
Let $C$ be a pricing function, and suppose $\widetilde{C}$ is a positive function satisfying the condition
\begin{equation}
\log\frac{1}{C(K)}\sim\log\frac{1}{\widetilde{C}(K)}
\label{E:ptich}
\end{equation}
as $K\rightarrow\infty$. Then
\begin{equation}
I(K)\sim\frac{\sqrt{2}}{\sqrt{T}}\left[\sqrt{\log K+\log\frac{1}{\widetilde{C}(K)}}
-\sqrt{\log\frac{1}{\widetilde{C}(K)}}\right] 
\label{E:leeo1s}
\end{equation}
as $K\rightarrow\infty$. 
\end{corollary}

\it Proof. \rm It follows from (\ref{E:repli}) that
\begin{equation}
I(K)\sim\frac{\sqrt{2}}{\sqrt{T}}\left[\sqrt{\log K+\log\frac{1}{\widetilde{C}(K)}}
-\sqrt{\log\frac{1}{\widetilde{C}(K)}}\right]\Lambda(K)
\label{E:lulu1}
\end{equation}
where
$$
\Lambda(K)=\frac{\sqrt{\log K+\log\frac{1}{\widetilde{C}(K)}}
+\sqrt{\log\frac{1}{\widetilde{C}(K)}}}{\sqrt{\log K+\log\frac{1}{C(K)}}
+\sqrt{\log\frac{1}{C(K)}}}.
$$

Our next goal is to prove that 
$\Lambda(K)\rightarrow 1$ as $K\rightarrow\infty$. We have
$$
\Lambda(K)=\frac{\sqrt{\Lambda_1(K)+\Lambda_2(K)}+\sqrt{\Lambda_2(K)}}{\sqrt{\Lambda_1(K)+1}+1}
$$
where
$$
\Lambda_1(K)=\frac{\log K}{\log\frac{1}{C(K)}}\quad\mbox{and}\quad\Lambda_2(K)
=\frac{\log\frac{1}{\widetilde{C}(K)}}{\log\frac{1}{C(K)}}.
$$
It is not hard to show that for all positive numbers $a$ and $b$, $|\sqrt{a+b}-\sqrt{a+1}|\le |\sqrt{b}-1|$. Therefore,
\begin{align}
\left|\Lambda(K)-1\right|&=\frac{\left|\sqrt{\Lambda_1(K)+\Lambda_2(K)}-\sqrt{\Lambda_1(K)+1}\right|
+\left|\sqrt{\Lambda_2(K)}-1\right|}{\sqrt{\Lambda_1(K)+1}+1} \nonumber \\
&\le\left|\sqrt{\Lambda_2(K)}-1\right|
\label{E:ptichk}
\end{align}
for $K> K_0$. It follows from (\ref{E:ptich}) and (\ref{E:ptichk}) that $\Lambda(K)\rightarrow 1$ as $K\rightarrow\infty$.
Next using (\ref{E:lulu1}) we see that (\ref{E:leeo1s}) holds. 

This completes the proof of Corollary \ref{C:geg}.
\section{Asymptotic behavior of the implied volatility as $K\rightarrow 0$}\label{S:kdo}
In this section, we turn out attention to the behavior of the implied volatility as the strike price tends to zero. 
We will first discuss a certain symmetry condition satisfied by the implied volatility. A similar condition can be found in Section 4 of \cite{keyL2}
(see also formula 2.9 in \cite{keyBFL}).

Let $C$ be a pricing function, and let $X$ be a corresponding stock price process. This process is defined on a filtered 
probability space 
$\left(\Omega,{\cal F},{\cal F}_t,\mathbb{P}^{*}\right)$, where $\mathbb{P}^{*}$ is a risk-free probability measure. 
As before, we assume that the interest rate $r$, the initial condition $x_0$, and the expiry $T$ are fixed, and denote by $\mu_T$ the 
distribution of the random variable $X_T$. The Black-Scholes pricing function $C_{BS}$ satisfies the following condition:
\begin{equation}
C_{BS}(T,K,\sigma)=x_0-Ke^{-rT}+\frac{Ke^{-rT}}{x_0}C_{BS}\left(T,\left(x_0e^{rT}\right)^2K^{-1},\sigma\right).
\label{E:atzero1}
\end{equation}
A similar formula holds for the pricing function $C$. Indeed, it is not hard to prove using the put/call parity condition that
\begin{align}
C(T,K)=x_0-Ke^{-rT}+\frac{Ke^{-rT}}{x_0}G\left(T,\left(x_0e^{rT}\right)^2K^{-1}\right),
\label{E:atzero5}
\end{align}
where the function $G$ is given by
\begin{equation}
G(T,K)=\frac{K}{x_0e^{rT}}P\left(T,\left(x_0e^{rT}\right)^2K^{-1}\right).
\label{E:em}
\end{equation}
It follows from (\ref{E:atzero5}) that
\begin{equation}
G\left(T,\left(x_0e^{rT}\right)^2K^{-1}\right)=x_0\int_0^Kd\mu_T(x)-\frac{x_0}{K}\int_0^Kxd\mu_T(x).
\label{E:emu}
\end{equation}

Define a family of Borel measures $\left\{\widetilde{\mu}_t\right\}_{t\ge 0}$ on $(0,\infty)$ as follows: 
For any Borel set $A$ in $(0,\infty)$, put
\begin{equation}
\widetilde{\mu}_t(A)=\frac{1}{x_0e^{rt}}\int_{\eta^{-1}(A)}xd\mu_t(x),
\label{E:nov}
\end{equation}
where $\displaystyle{\eta(K)=\left(x_0e^{rT}\right)^2K^{-1}}$, $K> 0$. It is not hard to see that $\widetilde{\mu}_t((0,\infty))=1$ for all $t\ge 0$. Moreover, for every 
Borel set $A$ in $(0,\infty)$, we have 
\begin{equation}
\int_{\eta(A)}d\widetilde{\mu}(x)=\frac{1}{x_0e^{rT}}\int_Axd\mu_T(x)\quad\mbox{and}\quad
\int_{\eta(A)}xd\widetilde{\mu}_T(x)=x_0e^{rT}\int_Ad\mu_T(x).
\label{E:atzero4}
\end{equation}
It follows from (\ref{E:em}) and (\ref{E:atzero4}) that
\begin{equation}
G(T,K)=e^{-rT}\int_K^{\infty}xd\widetilde{\mu}_T(x)-e^{-rT}K\int_K^{\infty}d\widetilde{\mu}_T(x).
\label{E:atzeroi}
\end{equation}
\begin{remark}\label{R:lala} \rm
Suppose that for every $t> 0$, the measure $\mu_t$ is absolutely continuous with respect to the Lebesgue measure on $(0,\infty)$. Denote the Radon-Nikodym derivative by $D_t$. 
Then for every
$t> 0$, the measure $\widetilde{\mu}_t$ admits a density $\widetilde{D}_t$ given by
$$
\widetilde{D}_t(x)=\left(x_0e^{rt}\right)^3x^{-3}D_t\left(\left(x_0e^{rt}\right)^2x^{-1}\right),\quad x> 0.
$$
\end{remark}

The next lemma provides a link between the asymptotic behavior of the implied 
volatility near infinity and near zero.
\begin{lemma}\label{L:link}
Let $C$ be a call pricing function and let $P$ be the corresponding put pricing function. Denote by $\left\{\mu_t\right\}_{t\ge 0}$ the family of 
marginal distributions of the stock price process $X$, and define a family of measures
$\left\{\widetilde{\mu}_t\right\}_{t\ge 0}$ by formula (\ref{E:nov}) and a function $G$ by formula (\ref{E:atzeroi}). Then $G$ is a call pricing function with the same interest rate $r$ 
and the initial condition $x_0$ as the pricing function $C$,
and with a stock price process $\widetilde{X}$ 
having $\left\{\widetilde{\mu}_t\right\}_{t\ge 0}$ 
as the family of its marginal distributions. 
\end{lemma}

\it Proof. \rm It suffices to prove that Conditions 1-5 in the characterization of call pricing functions formulated in the introduction
are valid for the function $G$. We have $\widetilde{\mu}_T([0,\infty))=1$. This equality follows from (\ref{E:atzero4}) and (\ref{E:mapr}). 
In addition, equality (\ref{E:vo}) holds for $\widetilde{\mu}_t$, by (\ref{E:atzero4}). Put
$$
V(T,K)=\int_K^{\infty}xd\widetilde{\mu}_T(x)-K\int_K^{\infty}d\widetilde{\mu}_T(x).
$$
Then $G(T,K)=e^{-rT}V(T,K)$. Moreover, the function $K\mapsto V(T,K)$ is convex on $[0,\infty)$ since its second distributional derivative coincides with the measure $\widetilde{\mu}_T$. This 
establishes conditions 1 and 2. The equality $G(0,K)=\left(x_0-K\right)_{+}$ can be obtained using (\ref{E:atzero4}) and (\ref{E:atzeroi}). 
This gives Condition 4. Next, we see that (\ref{E:atzeroi}) implies
$$
G(T,K)\le e^{-rT}\int_K^{\infty}xd\widetilde{\mu}_T(x),
$$
and hence
$\displaystyle{\lim_{K\rightarrow\infty}G(T,K)=0}$. This establishes Condition 5. In order to prove Condition 3 for $G$, we notice that (\ref{E:atzero5}) gives the following:
$$
G\left(T,e^{rT}K\right)=\frac{K}{x_0}C\left(T,e^{rT}\frac{x_0^2}{K}\right)+x_0-K.
$$
Now it is clear that Condition 3 for $G$ follows from the same condition for $C$. 
Therefore, $G$ is a call pricing function. 

This completes the proof of Lemma \ref{L:link}.

Let us denote by $I_C$ the implied volatility associated with the call pricing function $C$ and by $I_G$ the implied volatility associated with the call pricing function $G$.
Replacing $K$ by $I_C(K)$ in (\ref{E:atzero1}) and taking into account the equality $C_{BS}\left(K,I_C(K)\right)=C(K)$ and (\ref{E:atzero5}), we see that 
$$
C_{BS}\left(T,\left(x_0e^{rt}\right)^2K^{-1},I_C(T,K)\right)=G\left(T,\left(x_0e^{rT}\right)^2K^{-1}\right).
$$
Therefore, the following lemma holds:
\begin{lemma}\label{L:lala}
Let $C$ be a call pricing function and let $G$ be the call pricing function defined by (\ref{E:atzeroi}). Then 
\begin{equation}
I_C(T,K)=I_G\left(T,\left(x_0e^{rT}\right)^2K^{-1}\right)
\label{E:atzero7}
\end{equation}
for all $T\ge 0$ and $K> 0$.
\end{lemma}

Formula (\ref{E:atzero7}) can be interpreted as a symmetry condition for the implied volatility. For a certain class of
uncorrelated stochastic volatility models, a similar condition was established in \cite{keyRT} (see also \cite{keyGS3}). For the models from this class, we have 
$\displaystyle{I(T,K)=I\left(T,\left(x_0e^{rT}\right)^2K^{-1}\right)}$. 

Lemma \ref{L:lala} and the results obtained in Section \ref{S:BS} can be used to find sharp asymptotic formulas for the implied volatility as $K\rightarrow 0$. 
\begin{theorem}\label{T:genero}
Let $C$ be a call pricing function, and let $P$ be the corresponding put pricing function. Suppose that
$\widetilde{P}$ is a function such that 
\begin{equation}
P(K)\approx\widetilde{P}(K)\quad\mbox{as}\quad K\rightarrow 0.
\label{E:pp1}
\end{equation} 
Then the following asymptotic formula holds: 
\begin{equation}
I(K)=\frac{\sqrt{2}}{\sqrt{T}}\left[\sqrt{\log\frac{1}{\widetilde{P}(K)}}-\sqrt{\log\frac{K}{\widetilde{P}(K)}}\right]
+O\left(\left(\log\frac{K}{\widetilde{P}(K)}\right)^{-\frac{1}{2}}\log\log\frac{K}{\widetilde{P}(K)}\right)
\label{E:ppo1}
\end{equation}
as $K\rightarrow 0$. 
\end{theorem}

An important special case of Theorem \ref{T:genero} is as follows:
\begin{theorem}\label{T:cp}
Let $C$ be a call pricing function, and let $P$ be the corresponding put pricing function. Then
\begin{equation}
I(K)=\frac{\sqrt{2}}{\sqrt{T}}\left[\sqrt{\log\frac{1}{P(K)}}-\sqrt{\log\frac{K}{P(K)}}\right]
+O\left(\left(\log\frac{K}{P(K)}\right)^{-\frac{1}{2}}\log\log\frac{K}{P(K)}\right)
\label{E:ppo2}
\end{equation}
as $K\rightarrow 0$. 
\end{theorem}

\it Proof of Theorem \ref{T:genero}. \rm Formulas (\ref{E:pp1}) and (\ref{E:em}) imply that $G(K)\approx\widetilde{G}(K)$ as $K\rightarrow\infty$ where
\begin{equation}
\widetilde{G}(K)=K\widetilde{P}\left(\left(x_0e^{rT}\right)^2K^{-1}\right).
\label{E:pp2}
\end{equation}
Applying Theorem \ref{T:genero} to $G$ and $\widetilde{G}$, we get
\begin{equation}
I_G(K)=\frac{\sqrt{2}}{\sqrt{T}}\left[\sqrt{\log K+\log\frac{1}{\widetilde{G}(K)}}-\sqrt{\log\frac{1}{\widetilde{G}(K)}}\right]
+O\left(\left(\log\frac{1}{\widetilde{G}(K)}\right)^{-\frac{1}{2}}\log\log\frac{1}{\widetilde{G}(K)}\right)
\label{E:pp3}
\end{equation}
as $K\rightarrow\infty$. It follows from (\ref{E:pp3}), (\ref{E:atzero7}), and (\ref{E:pp2}) that
\begin{align}
I_C(K)&=\frac{\sqrt{2}}{\sqrt{T}}\left[\sqrt{\log\frac{\left(x_0e^{rT}\right)^2}{K}+\log\frac{K}{\left(x_0e^{rT}\right)^2\widetilde{P}(K)}}
-\sqrt{\log\frac{K}{\left(x_0e^{rT}\right)^2\widetilde{P}(K)}}\right] \nonumber \\
&\quad+O\left(\left(\log\frac{K}{\widetilde{P}(K)}\right)^{-\frac{1}{2}}\log\log\frac{K}{\widetilde{P}(K)}\right)
\label{E:pp4}
\end{align}
as $K\rightarrow 0$. It is not hard to see that formula (\ref{E:pp2}) implies 
\begin{equation}
K\left[\widetilde{P}(K)\right]^{-1}\rightarrow\infty\quad\mbox{as}\quad K\rightarrow 0.
\label{E:bu}
\end{equation}
Finally, using (\ref{E:pp4}) and the mean value theorem, we get (\ref{E:ppo1}).

This completes the proof of Theorem \ref{T:genero}.
\begin{remark}\label{R:ge} \rm Note that (\ref{E:bu}) shows that
$K[P(K)]^{-1}\rightarrow\infty$ as $K\rightarrow 0$.
\end{remark}
\section{Sharp asymptotic formulas for the implied volatility and Lee's moment formulas}\label{S:lee}
In \cite{keyL2}, Roger Lee obtained important asymptotic formulas for the implied volatility. We will next formulate Lee's results.
\begin{theorem}\label{T:le1}
Let $I$ be the implied volatility associated with a call pricing function $C$. Define a number $\tilde{p}$ by
\begin{equation}
\tilde{p}=\sup\left\{p\ge 0:\mathbb{E}^{*}\left[X_T^{1+p}\right]<\infty\right\}.
\label{E:p}
\end{equation}
Then the following equality holds:
\begin{equation}
\limsup_{K\rightarrow\infty}\frac{TI(K)^2}{\log K}=\psi(\tilde{p})
\label{E:leef1}
\end{equation}
where the function $\psi$ is given by
\begin{equation}
\psi(u)=2-4\left(\sqrt{u^2+u}-u\right),\quad u\ge 0.
\label{E:pps}
\end{equation}
\end{theorem}
\begin{theorem}\label{T:le2}
Let $I$ be the implied volatility associated with a call pricing function $C$. Define a number $\tilde{q}$ by
\begin{equation}
\tilde{q}=sup\left\{q\ge 0:\,\mathbb{E}\left[X_T^{-q}\right]<\infty\right\}.
\label{E:pmf2}
\end{equation}
Then the following formula holds:
\begin{equation}
\limsup_{K\rightarrow 0}\frac{TI(K)^2}{\log\frac{1}{K}}=\psi(\tilde{q}).
\label{E:pmf1}
\end{equation}
\end{theorem}
 
Formulas (\ref{E:leef1}) and (\ref{E:pmf1}) are called Lee's moment formulas, and the numbers $1+\tilde{p}$ and $\tilde{q}$ are called the right-tail index and the left-tail index 
of the distribution of the stock price $X_T$, respectively. These numbers show how fast the tails of the distribution of the stock price decay. 

We will next show how to derive Lee's moment formula (\ref{E:leef1}) using our formula (\ref{E:repli}). In order to see how (\ref{E:repli})
is linked to Lee's formulas, we note that for every $a>0$,
\begin{equation}
\sqrt{1+a}-\sqrt{a}=\left(1-2\left(\sqrt{a^2+a}-a\right)\right)^{\frac{1}{2}}=\sqrt{2^{-1}\psi(a)}.
\label{E:fo1}
\end{equation}
Therefore, Lee's formulas (\ref{E:leef1}) and (\ref{E:pmf1}) can be rewritten as follows:
\begin{equation}
\limsup_{K\rightarrow\infty}\frac{\sqrt{T}I(K)}{\sqrt{2\log K}}=\sqrt{1+\tilde{p}}-\sqrt{\tilde{p}}
\label{E:leef1r}
\end{equation}
and
\begin{equation}
\limsup_{K\rightarrow 0}\frac{\sqrt{T}I(K)}{\sqrt{2\log\frac{1}{K}}}=\sqrt{1+\tilde{q}}-\sqrt{\tilde{q}}.
\label{E:pmf1s}
\end{equation}

Our next goal is to establish formula (\ref{E:leef1r}). 
\begin{lemma}\label{L:fol1}
Let $C$ be a call pricing function and put 
\begin{equation}
l=\liminf_{K\rightarrow\infty}\,(\log K)^{-1}\log\frac{1}{C(K)}.
\label{E:leef5}
\end{equation}
Then 
\begin{equation}
\limsup_{K\rightarrow\infty}\frac{\sqrt{T}I(K)}{\sqrt{2\log K}}=\sqrt{1+l}-\sqrt{l}.
\label{E:san}
\end{equation}
\end{lemma}

\it Proof of Lemma \ref{L:fol1}. \rm Observe that (\ref{E:repli}) implies
\begin{align}
\frac{\sqrt{T}I(K)}{\sqrt{2\log K}}&=\sqrt{1+\frac{\log\frac{1}{C(K)}}{\log K}}-\sqrt{\frac{\log\frac{1}{C(K)}}{\log K}}
+O\left((\log K)^{-\frac{1}{2}}\left(\log\frac{1}{C(K)}\right)^{-\frac{1}{2}}\log\log\frac{1}{C(K)}\right) \nonumber \\
&=\left[\sqrt{1+\frac{\log\frac{1}{C(K)}}{\log K}}+\sqrt{\frac{\log\frac{1}{C(K)}}{\log K}}\right]^{-1}
+O\left((\log K)^{-\frac{1}{2}}\left(\log\frac{1}{C(K)}\right)^{-\frac{1}{2}}\log\log\frac{1}{C(K)}\right)
\label{E:sam}
\end{align}
as $K\rightarrow\infty$. It is clear that (\ref{E:san}) follows from (\ref{E:sam}).

Let us continue the proof of formula (\ref{E:leef1r}). Denote by $\rho_T$ the complementary distribution function of $X_T$ given by
$\rho_T(y)=\mathbb{P}\left[X_T> y\right]$, $y>0$. Then
\begin{equation}
C(K)=e^{-rT}\int_K^{\infty}\rho_T(y)dy,\quad K> 0.
\label{E:leef2}
\end{equation}
Define the following numbers:
\begin{equation}
r^{*}=\sup\left\{r\ge 0:C(K)=O\left(K^{-r}\right)\quad\mbox{as}\quad K\rightarrow\infty\right\},
\label{E:leef3}
\end{equation}
and
\begin{equation}
s^{*}=\sup\left\{s\ge 0:\rho_T(y)=O\left(y^{-(1+s)}\right)\quad\mbox{as}\quad y\rightarrow\infty\right\}.
\label{E:leef4}
\end{equation}
\begin{lemma}\label{L:lee}
The numbers $\tilde{p}$, $l$, $r^{*}$, and $s^{*}$ given by (\ref{E:p}), (\ref{E:leef5}), (\ref{E:leef3}), 
and (\ref{E:leef4}), respectively, are all equal.
\end{lemma}

\it Proof. \rm If $s^{*}=0$, then the inequality $s^{*}\le r^{*}$ is trivial. If $s> 0$ is such that 
$\rho_T(y)=O\left(y^{-(1+s)}\right)$ as $y\rightarrow\infty$, then
$$
C(K)=O\left(\int_K^{\infty}y^{-(1+s)}dy\right)=O\left(K^{-s}\right)
$$
as $K\rightarrow\infty$. Hence $s^{*}\le r^{*}$.

Next let $r\ge 0$ be such that $C(K)=O\left(K^{-r}\right)$ as $K\rightarrow\infty$. Then (\ref{E:leef2}) shows that
there exists $c> 0$ for which
$$
cK^{-r}\ge e^{-rT}\int_K^{\infty}\rho_T(y)dy\ge e^{-rT}\int_K^{2K}\rho_T(y)dy\ge e^{-rT}\rho_T(2K)K,\quad K> K_0.
$$
Therefore, $\rho_T(K)=O\left(K^{-(r+1)}\right)$ as $K\rightarrow\infty$. It follows that $r^{*}\le s^{*}$.
This proves the equality $r^{*}=s^{*}$.

Suppose that $0< l<\infty$. Then for every $\varepsilon> 0$, there exists $K_{\varepsilon}> 0$ such that
$$
\log\frac{1}{C(K)}\ge(l-\varepsilon)\log K,\quad K> K_{\varepsilon}.
$$
Therefore $C(K)\le K^{-l+\varepsilon},\quad K> K_{\varepsilon}$. It follows that $l-\varepsilon\le r^{*}$ 
for all $\varepsilon> 0$, and hence $l\le r^{*}$.
The inequality $l\le r^{*}$ also holds if $l=0$ or $l=\infty$. This fact can be established similarly.

To prove the inequality $r^{*}\le l$, suppose that $r^{*}\neq 0$ and $r< r^{*}$. Then $C(K)=O\left(K^{-r}\right)$ 
as $K\rightarrow\infty$, and hence $\displaystyle{\frac{1}{C(K)}\ge cK^r}$
for some $c> 0$ and all $K> K_0$. It follows that $\displaystyle{\log\frac{1}{C(K)}\ge\log c+r\log K,\quad K> K_0}$ and
$$
\frac{\log\frac{1}{C(K)}}{\log K}\ge\frac{\log c}{\log K}+r.
$$
Now it is clear that 
\begin{equation}
\liminf_{K\rightarrow\infty}\frac{\log\frac{1}{C(K)}}{\log K}\ge r.
\label{E:leef5}
\end{equation}
Using (\ref{E:leef5}), we see that $l\ge r^{*}$. If $r^{*}=0$, then the inequality $l\ge r^{*}$ is trivial. This proves that $l=r^{*}=s^{*}$.

It is clear that for all $p\ge 0$,
\begin{equation}
\mathbb{E}\left[X_T^{1+p}\right]=(1+p)\int_0^{\infty}y^p\rho_T(y)dy.
\label{E:leef6}
\end{equation}
Suppose that $s^{*}=0$, then the inequality $s^{*}\le\tilde{p}$ is trivial. If for some $s> 0$, $\rho_T(y)=O\left(y^{-(1+s)}\right)$ as $y\rightarrow\infty$, 
then it is not hard to see using (\ref{E:leef6}) that $\mathbb{E}\left[X_T^{1+p}\right]<\infty$ 
for all $p< s$. It follows that $s^{*}\le\tilde{p}$.

On the other hand, if $\mathbb{E}\left[X_T^{1+p}\right]<\infty$ for some $p\ge 0$, then
\begin{equation}
M> \int_x^{\infty}y^p\rho_T(y)dy\ge K^{p}\int_K^{\infty}\rho_T(y)dy=e^{rT}K^pC(K).
\label{E:leef7}
\end{equation}
In the proof of (\ref{E:leef7}), we used (\ref{E:leef6}) and (\ref{E:leef2}). 
It follows from (\ref{E:leef7}) that $C(K)=O\left(K^{-p}\right)$ as $K\rightarrow\infty$,
and hence $\tilde{p}\le r^{*}$. 

This completes the proof of Lemma \ref{L:lee}.

To finish the proof of formula (\ref{E:leef1}), we observe that (\ref{E:san}) and the equality $l=\tilde{p}$ 
in Lemma \ref{L:lee} imply formula (\ref{E:leef1r}).

It will be explained next how to obtain formula (\ref{E:pmf1s}) from formula (\ref{E:ppo2}). 
Taking into account (\ref{E:ppo2}), we get the following lemma:
\begin{lemma}\label{L:left}
Let $C$ be a call pricing function and define a number by
\begin{equation}
m=\liminf_{K\rightarrow 0}\left(\log\frac{1}{K}\right)^{-1}\log\frac{1}{P(K)}.
\label{E:pmf4}
\end{equation}
Then
\begin{equation}
\limsup_{K\rightarrow 0}\frac{\sqrt{T}I(K)}{\sqrt{2\log\frac{1}{K}}}=\sqrt{m}-\sqrt{m-1}.
\label{E:pmf3}
\end{equation}
\end{lemma}
 
The inequality $m\ge 1$ where $m$ is defined by (\ref{E:pmf4}) follows from Remark \ref{R:ge}. Put 
$$
\eta_T(y)=\mathbb{P}\left[X_T\le y\right]=1-\rho_T(y),\quad y\ge 0.
$$ 
Then 
$$
P(K)=e^{-rT}\int_0^K\eta_T(y)dy
$$
and
$$
\mathbb{E}\left[X_T^{-q}\right]=q\int_0^{\infty}y^{-q-1}\eta_T(y)dy
$$
for all $q> 0$. Note that $\eta_T(0)=\mathbb{P}\left[X_T=0\right]$.

Consider the following numbers:
\begin{equation}
u^{*}=\sup\left\{u\ge 1:\, P(K)=O\left(K^{u}\right)\quad\mbox{as}\quad K\rightarrow 0\right\}
\label{E:pmf5}
\end{equation}
and
\begin{equation}
v^{*}=\sup\left\{v\ge 0:\, \eta_T(y)=O\left(y^v\right)\quad\mbox{as}\quad y\rightarrow 0\right\}.
\label{E:pmf6}
\end{equation}
It is not hard to see, using the same ideas as in the proof of Lemma \ref{L:lee}, that the following lemma holds:
\begin{lemma}\label{L:ree}
The numbers $\tilde{q}$, $m$, $u^{*}$, and $v^{*}$ defined by (\ref{E:pmf2}), (\ref{E:pmf4}), 
(\ref{E:pmf5}), and (\ref{E:pmf6}), respectively, satisfy the condition
$\tilde{q}+1=m=u^{*}=v^{*}+1$.
\end{lemma}

Now it is clear that formula (\ref{E:pmf1s}) follows form (\ref{E:pmf3}) and Lemma \ref{L:ree}.
\section{Sharp asymptotic formulas for the implied volatility and the tail-wing formulas of Benaim and Friz}\label{S:cBF}
In \cite{keyBF}, Benaim and Friz studied asymptotic relations between a given call pricing function, the implied volatility associated with it, 
and the law of the stock returns, under an additional 
assumption that there exist non-trivial moments of the stock price. We will next give
several definitions from the theory of regularly varying functions (these definitions will be needed in the remaining part of the paper), and then formulate 
some of the results obtained in \cite{keyBF}. 
\begin{definition}\label{D:rvf}
Let $\alpha\in\mathbb{R}$ and let $f$ be a Lebesgue measurable function defined on some neighborhood of infinity. 
The function $f$ is called regularly varying with index $\alpha$ if the following condition holds: For every $\lambda> 0$, 
$\displaystyle{\frac{f(\lambda x)}{f(x)}\rightarrow 
\lambda^{\alpha}}$ as $x\rightarrow\infty$. The class consisting of all regularly varying functions with index $\alpha$ is denoted by $R_{\alpha}$. 
Functions belonging to the class $R_0$ are called slowly varying.
\end{definition}

The following asymptotic formula is valid for all functions $f\in R_{\alpha}$ with $\alpha> 0$:
\begin{equation}
-\log\int_K^{\infty}e^{-f(y)}dy\sim f(K)\quad\mbox{as}\quad K\rightarrow\infty.
\label{E:bing}
\end{equation}
(see Theorem 4.12.10 (i) in \cite{keyBGT}). This result is known as Bingham's Lemma.  
\begin{definition}\label{D:svf}
Let $\alpha\in\mathbb{R}$ and let $f$ be a positive function defined on some neighborhood of infinity. 
The function $f$ is caled smoothly varying with index $\alpha$
if the function $h(x)=\log f\left(e^x\right)$ is infinitely differentiable and $h^{\prime}(x)\rightarrow\alpha$, $h^{(n)}(x)\rightarrow 0$ for all integers $n\ge 2$ 
as $x\rightarrow\infty$.
\end{definition}

An equivalent definition of the class $SR_{\alpha}$ is as follows:
\begin{equation}
f\in SR_{\alpha}\Leftrightarrow\lim_{x\rightarrow\infty}\frac{x^nf^{(n)}(x)}{f(x)}=\alpha(\alpha-1)\ldots(\alpha-n+1)
\label{E:fact}
\end{equation}
for all $n\ge 1$.
\begin{definition}\label{D:svr}
Let $g$ be a function on $(0,\infty)$ such that $g(x)\downarrow 0$ as $g\rightarrow\infty$. A function $l$ defined on $(a,\infty)$, $a\ge 0$, is called slowly varying with remainder
$g$ if $l\in R_0$ and $\displaystyle{\frac{l(\lambda x)}{l(x)}-1=O(g(x))}$ as $x\rightarrow\infty$ for all $\lambda> 1$.
\end{definition}
Definitions (\ref{D:rvf}) - (\ref{D:svr}) can be found in \cite{keyBGT}. The theory of regularly varying functions has interesitng applications in financial mathematics. 
Besides \cite{keyBF}, \cite{keyBF1}, and \cite{keyBFL}, such functions appear in the study 
of Pareto-type tails of distributions of stock returns (see, e.g, \cite{keyC} and the references therein). We refer the reader to \cite{keyBGTS} for more 
information on Pareto-type distributions and their applications. Pareto-type distributions are defined as follows. Let $X$ be a random variable on a probability 
space $(\Omega,{\cal F},\mathbb{P})$, 
and let $F$ be the distribution function of $X$ given by $F(y)=\mathbb{P}[X\le y]$, $y\in\mathbb{R}$. By $\bar{F}$ 
we denote the complementary distribution function of $X$, 
that is, the function 
$\bar{F}(y)=1-F(y),\quad y\in\mathbb{R}$.
The distribution $F$ is called a Pareto-type distribution if the complementary 
distribution function $\bar{F}$ is well fit by a power law. More precisely, $F$ is a Pareto-type distribution with index $\alpha> 0$ iff
there exists a function $h\in R_0$ such that 
$\bar{F}(y)\sim y^{-\alpha}h(y)\quad\mbox{as}\quad y\rightarrow\infty$. 

We will next formulate some of the results obtained in \cite{keyBF} adapting them to our notation (see Theorem 1 in \cite{keyBF}). 
Note that Benaim and Friz use a different normalization in the Black-Scholes formula and consider the normalized 
implied volatility as a function of the log-strike $k$. In the formulation of Theorem \ref{T:befr1} below, the function $\psi$ defined by
$\psi(u)=2-4\left(\sqrt{u^2+u}-u\right)$ is used. This function has already appeared in Section \ref{S:lee}.
It is clear that $\psi$ is strictly decreasing on the interval $[0,\infty]$ and maps 
this interval onto the interval $[0,2]$. 
\begin{theorem}\label{T:befr1}
Let $C$ be a pricing function and suppose that
\begin{equation}
\mathbb{E}^{*}\left[X_T^{1+\varepsilon}\right]<\infty\quad\mbox{for some}\quad \varepsilon> 0.
\label{E:qu1}
\end{equation}
Then the following statements hold:
\begin{enumerate}
\item If $C(K)=\exp\left\{-\eta(\log K)\right\}$ with $\eta\in R_{\alpha}$, $\alpha> 0$, 
then 
\begin{equation}
I(K)\sim\frac{\sqrt{\log K}}{\sqrt{T}}\sqrt{\psi\left(-\frac{\log C(K)}{\log K}\right)}\quad\mbox{as}\quad K\rightarrow\infty.
\label{E:vost1}
\end{equation}
\item If $\bar{F}(y)=\exp\left\{-\rho(\log y)\right\}$ with $\rho\in R_{\alpha}$, $\alpha> 0$, then
\begin{equation}
I(K)\sim\frac{\sqrt{\log K}}{\sqrt{T}}\sqrt{\psi\left(-\frac{\log[K\rho(K)]}{\log K}\right)}\quad\mbox{as}\quad K\rightarrow\infty.
\label{E:vosto1}
\end{equation}
\item If the distribution $\mu_T$ of the stock price $X_T$ admits a density $D_T$ and if 
\begin{equation}
D_T(x)=\frac{1}{x}\exp\left\{-h(\log x)\right\}
\label{E:soo}
\end{equation}
as $x\rightarrow\infty$, where $h\in R_{\alpha}$, $\alpha> 0$, then
\begin{equation}
I(K)\sim\frac{\sqrt{\log K}}{\sqrt{T}}\sqrt{\psi\left(-\frac{\log[K^2D_T(K)]}{\log K}\right)}\quad\mbox{as}\quad K\rightarrow\infty.
\label{E:vostok1}
\end{equation}
\end{enumerate} 
\end{theorem}

Note that $V(k)$ and $c(k)$ in \cite{keyBF} correspond in our notation to 
$\sqrt{T}I(K)$ and  $e^{rT}C(K)$, respectively. 
We also take into account that the distribution density of the stock returns $f(k)$, where $k$ stands for the log-strike, is related to the density $D_T$ by the formula 
$f(k)=e^kD_T(e^k)$.

The formulas contained in Theorem \ref{T:befr1} are called the right-tail-wing formulas of Benaim and Friz. The idea to express the asymptotic properties of the implied volatility in terms of the 
behavior of the distribution density of the stock price has also been exploited in \cite{keyGS2} and \cite{keyGS4} in the case of special stock price models with stochastic volatility. 

Our next goal is to derive Theorem \ref{T:befr1} from Corollary \ref{C:geg}.
The next statement is nothing else but this corollary in disguise. 
\begin{corollary}\label{C:pob}
For any pricing function $C$,
\begin{equation}
I(K)=\frac{\sqrt{\log K}}{\sqrt{T}}\sqrt{\psi\left(-\frac{\log C(K)}{\log K}\right)}
+O\left(\left(\log\frac{1}{C(K)}\right)^{-\frac{1}{2}}\log\log\frac{1}{C(K)}\right)
\label{E:vost2}
\end{equation}
as $K\rightarrow\infty$, where $\psi$ is defined by (\ref{E:pps}).
\end{corollary}

The equivalence of formulas (\ref{E:repli}) and (\ref{E:vost2}) can be easily shown using (\ref{E:fo1}).
\begin{remark}\label{R:stro} \rm
It follows from Corollary \ref{C:pob} that formula (\ref{E:vost1}) holds for any call pricing function, and hence no restrictions are needed in Part 1
of Theorem \ref{T:befr1}. Moreover, formula (\ref{E:vost2}) contains an error term, which is absent in formula (\ref{E:vost1}). 
\end{remark}

We will next brifly explain how to obtain (\ref{E:vosto1}) and (\ref{E:soo}). We will prove a slightly more general statement assuming that 
\begin{equation}
\bar{F}(y)\approx\exp\left\{-\rho(\log y)\right\}
\label{E:mor1}
\end{equation} 
as $y\rightarrow\infty$ in Part 2 of Theorem \ref{T:befr1} and
\begin{equation}
D_T(x)\approx x^{-1}\exp\left\{-h(\log x)\right\}
\label{E:mor2}
\end{equation} 
as $x\rightarrow\infty$ in Part 3.
Some of the ideas used in the proof below are borrowed from \cite{keyBF} (see the proofs in Section 3 of \cite{keyBF}). With no loss of generality, we may suppose that $\alpha\ge 1$. 
The proof of the tail-wing formulas is based on Bingham's Lemma and the following equalities:
\begin{equation}
C(K)=e^{-rT}\int_K^{\infty}\bar{F}(y)dy
\label{E:cummul}
\end{equation}
and 
\begin{equation}
\bar{F}(y)=\int_y^{\infty}D_T(x)dx.
\label{E:ddd}
\end{equation}
If $\alpha> 1$ in Parts 2 or 3 of Theorem \ref{T:befr1}, then the moment condition (\ref{E:qu1}) holds and we have $\rho(u)-u\in R_{\alpha}$ in Part 2 and $h(u)-u\in R_{\alpha}$ in Part 3.
If $\alpha=1$, then the moment condition gives $\rho(u)-u\in R_1$ in Part 2 and $h(u)-u\in R_1$ in Part 3 (see Section 3 in \cite{keyBF}). 

Suppose that (\ref{E:mor1}) holds. Put $\lambda(u)=\rho(u)-u$. Then we have
$C(K)\approx\widehat{C}(K)$ as $K\rightarrow\infty$,
where
$$
\widehat{C}(K)=\int_{\log K}^{\infty}\exp\left\{-\lambda(u)\right\}du.
$$
Applying formula (\ref{E:bing}) to the function $\lambda$, we obtain
$$
\log\frac{1}{\widehat{C}(K)}\sim\lambda(\log K)=\log\frac{1}{K\rho(K)}
$$
as $K\rightarrow\infty$. Since $C(K)\approx\widehat{C}(K)$, we also have 
$$
\log\frac{1}{C(K)}\sim\log\frac{1}{\widehat{C}(K)},
$$
and hence
$$
\log\frac{1}{C(K)}\sim\log\frac{1}{K\rho(K)}
$$
as $K\rightarrow\infty$. Now it clear that formula (\ref{E:vosto1}) follows from (\ref{E:leeo1s}) and (\ref{E:fo1}).

Next assume that equality (\ref{E:mor2}) holds. Then (\ref{E:ddd}) implies (\ref{E:mor1}) with
$$
\rho(y)=-\log\int_y^{\infty}e^{-h(u)}du.
$$
Applying Bingham's Lemma, we see that $\rho\in R_{\alpha}$. This reduces the case of the distribution density $D_T$ of the stock price in 
Theorem \ref{T:befr1} to that of the complement distribution function $\bar{F}$.
\begin{remark}\label{R:esh1} \rm
The tail-wing formula (\ref{E:vosto1}) also holds provided that $\alpha=1$ and $\rho(u)-u\in R_{\beta}$ with $0<\beta\le 1$. A similar statement is true in the case of formula (\ref{E:vostok1}). 
The proof of these assertions does not differ from the proof given above. 
Interesting examples here are $\rho(u)=u+u^{\beta}$ if $\beta< 1$ and $\displaystyle{\rho(u)=u+\frac{u}{\log u}}$ if $\beta=1$. Note that the moment condition does not hold in these cases.
\end{remark}

Formulas (\ref{E:vosto1}) and (\ref{E:vostok1}) do not include error estimates. Our next goal is to find asymptotic formulas for the implied volatility which contain error estimates. 
We can do it under certian smoothness assumptions on the functions $\rho$ and $h$ appearing in Theorem \ref{T:befr1}.
\begin{theorem}\label{T:strousa}
Let $C$ be a call pricing function and let $\bar{F}$ be the complementary distribution function of the stock price $X_T$.  Suppose that
\begin{equation}
\bar{F}(y)\approx\exp\left\{-\rho(\log y)\right\}
\label{E:vostik}
\end{equation} 
as $y\rightarrow\infty$, where $\rho$ is a function such that either $\rho\in SR_{\alpha}$ with $\alpha> 1$, or $\rho\in SR_1$ 
and $\lambda(u)=\rho(u)-u\in R_{\beta}$ for some $0<\beta\le 1$. Then 
\begin{equation}
I(K)=\frac{\sqrt{2}}{\sqrt{T}}\left(\sqrt{\rho(\log K)}-\sqrt{\rho(\log K)-\log K}\right)
+O\left(\frac{\log\left[\rho(\log K)\right]}{\sqrt{\rho(\log K)}}\right)
\label{E:vostrik}
\end{equation}
as $K\rightarrow\infty$. 
\end{theorem}
\begin{theorem}\label{T:stroussa}
Let $C$ be a call pricing function and let $D_T$ be the distribution density of the stock price $X_T$.
Suppose that
\begin{equation}
D_T(x)\approx\frac{1}{x}\exp\left\{-h(\log x)\right\}
\label{E:usl10}
\end{equation}
as $x\rightarrow\infty$, where $h$ is a function such that either $h\in SR_{\alpha}$ with $\alpha> 1$, 
or $h\in SR_1$ and $g(u)=h(u)-u\in SR_{\beta}$ for some $0<\beta\le 1$. Then 
\begin{equation}
I(K)=\frac{\sqrt{2}}{\sqrt{T}}\left(\sqrt{h(\log K)}-\sqrt{h(\log K)-\log K}\right)
+O\left(\frac{\log\left[h(\log K)\right]}{\sqrt{h(\log K)}}\right)
\label{E:vost5}
\end{equation}
as $K\rightarrow\infty$. 
\end{theorem}
\begin{remark}\label{R:bf} \rm Formulas (\ref{E:vostrik}) and (\ref{E:vost5}) are equivalent to the formulas
$$
I(K)=\frac{\sqrt{\log K}}{\sqrt{T}}\sqrt{\psi\left(\frac{\rho(\log K)-\log K}{\log K}\right)}
+O\left(\frac{\log\left[\rho(\log K)\right]}{\sqrt{\left(\rho(\log K)\right)}}\right),\quad K\rightarrow\infty,
$$
and
$$
I(K)=\frac{\sqrt{\log K}}{\sqrt{T}}\sqrt{\psi\left(\frac{h(\log K)-\log K}{\log K}\right)}
+O\left(\frac{\log\left[h(\log K)\right]}{\sqrt{\left(h(\log K)\right)}}\right),\quad K\rightarrow\infty,
$$
respectively, where the function $\psi$ is defined by (\ref{E:pps}). If equality holds in (\ref{E:vostik}) and (\ref{E:usl10}), then we get the following tail-wing formulas 
with error estimates:
$$
I(K)=\frac{\sqrt{\log K}}{\sqrt{T}}\sqrt{\psi\left(-\frac{\log\left[K\bar{F}(K)\right]}{\log K}\right)}
+O\left(\left(\log\frac{1}{\left[K\bar{F}(K)\right]}\right)^{-\frac{1}{2}}\log\log\frac{1}{\left[K\bar{F}(K)\right]}\right)
$$
and
$$
I(K)=\frac{\sqrt{\log K}}{\sqrt{T}}\sqrt{\psi\left(-\frac{\log\left[K^2D_T(K)\right]}{\log K}\right)}
+O\left(\left(\log\frac{1}{\left[K^2D_T(K)\right]}\right)^{-\frac{1}{2}}\log\log\frac{1}{\left[K^2D_T(K)\right]}\right)
$$
as $K\rightarrow\infty$. 
\end{remark}

We will next prove Theorem \ref{T:stroussa}. The proof of Theorem \ref{T:strousa} is similar, but less complicated. We leave it as an exercise for the reader.

\it Proof of Theorem \ref{T:stroussa}. \rm  
We borrow several ideas used in the proof of formula (\ref{E:bing}) (see Theorem 4.12.10 (i) in \cite{keyBGT}). The following lemma is standard:
\begin{lemma}\label{L:sa}
Suppose $r\in SR_{\alpha}$ with $\alpha> 0$. Then
$$
\int_x^{\infty}e^{-r(u)}du=\frac{e^{-r(x)}}{r^{\prime}(x)}\left(1+O\left(\frac{1}{r(x)}\right)\right)
$$
as $x\rightarrow\infty$.
\end{lemma}

\it Proof. \rm 
Using the integration by parts formula, we see that
\begin{equation}
\int_x^{\infty}e^{-r(u)}du=\frac{e^{-r(x)}}{r^{\prime}(x)}-\int_x^{\infty}e^{-r(u)}\rho_1(u)du
\label{E:sa4}
\end{equation}
where 
$$
\rho_1(u)=\left(\frac{1}{r^{\prime}(u)}\right)^{\prime}=\frac{r^{\prime\prime}(u)}{r^{\prime}(u)^2}.
$$
It follows from (\ref{E:fact}) that
$\displaystyle{\left|\rho_1(u)\right|=O\left(r(u)^{-1}\right)}$
as $u\rightarrow\infty$. Using (\ref{E:fact}) again, we obtain
\begin{align}
\int_x^{\infty}e^{-r(u)}\rho_1(u)du&=O\left(\int_x^{\infty}r^{\prime}(u)e^{-r(u)}\frac{1}{r(u)r^{\prime}(u)}du\right) \nonumber \\
&=O\left(\int_x^{\infty}r^{\prime}(u)e^{-r(u)}\frac{u}{r(u)^2}du\right)
\label{E:sa5}
\end{align}
as $x\rightarrow\infty$. Since for every $\varepsilon> 0$, the function $\displaystyle{u\mapsto e^{-\varepsilon r(u)}ur(u)^{-2}}$ is ultimately decreasing,
(\ref{E:sa5}) implies that
\begin{equation}
\int_x^{\infty}e^{-r(u)}\rho_1(u)du=O\left(e^{-r(x)}\frac{x}{r(x)^2}\right)
\label{E:sa6}
\end{equation}
as $x\rightarrow\infty$. Now Lemma \ref{L:sa} follows from (\ref{E:sa4}), (\ref{E:sa6}), and (\ref{E:fact}). 

Let us continue the proof of Theorem \ref{T:stroussa}. For $h\in SR_{\alpha}$ with $\alpha> 1$, we have $g\in SR_{\alpha}$. 
On the other hand, if $\alpha=1$, we assume that $g\in SR_{\beta}$ with $0<\beta\le 1$. Consider the following functions: 
$$
\widetilde{D}_T(x)=\frac{1}{x}\exp\left\{-h(\log x)\right\}\quad\mbox{and}\quad
\widehat{C}(K)=K^2\widetilde{D}_T(K)=\exp\left\{-g(\log K)\right\}.
$$
We have
\begin{equation}
C(K)\approx\int_{\log K}^{\infty}e^{-g(u)}du-K\int_{\log K}^{\infty}e^{-h(u)}du
\label{E:sss1}
\end{equation}
as $K\rightarrow\infty$. 
Now applying Lemma \ref{L:sa} we get
$$
\int_{\log K}^{\infty}e^{-g(u)}du=\frac{Ke^{-h(\log K)}}{g^{\prime}(\log K)}\left(1+O\left(\frac{1}{g(\log K)}\right)\right)
$$
and
$$
K\int_{\log K}^{\infty}e^{-h(u)}du=\frac{Ke^{-h(\log K)}}{h^{\prime}(\log K)}\left(1+O\left(\frac{1}{h(\log K)}\right)\right)
$$
as $K\rightarrow\infty$. It follows that
\begin{equation}
\int_{\log K}^{\infty}e^{-g(u)}du-K\int_{\log K}^{\infty}e^{-h(u)}du
=\frac{Ke^{-h(\log K)}}{h^{\prime}(\log K)g^{\prime}(\log K)}\left(1+O\left(\frac{h(\log K)}{g(\log K)\log K}\right)\right)
\label{E:sa7}
\end{equation}
as $K\rightarrow\infty$. In the proof of (\ref{E:sa7}) we used (\ref{E:fact}). Next employing (\ref{E:sss1}), (\ref{E:sa7}), and (\ref{E:fact}) we obtain
\begin{equation}
C(K)\approx\widetilde{C}(K)\quad\mbox{where}\quad  
\widetilde{C}(K)=\frac{K(\log K)^2e^{-h(\log K)}}{h(\log K)g(\log K)}.
\label{E:stro0}
\end{equation}
Next we see that
\begin{equation}
\log\frac{1}{\widetilde{C}(K)}-\log\frac{1}{\widehat{C}(K)}=\log\frac{h(\log K)g(\log K)}{(\log K)^2}.
\label{E:aia}
\end{equation}
It follows from (\ref{E:aia}) that there exists $a> 0$ such that
\begin{equation}
\left|\log\frac{1}{\widetilde{C}(K)}-\log\frac{1}{\widehat{C}(K)}\right|\le a\log\log\frac{1}{\widehat{C}(K)},\quad K>K_1.
\label{E:ist}
\end{equation}
Indeed, if $\alpha> 1$, we can take $a>\frac{2\alpha-2}{\alpha}$ in (\ref{E:ist}), and if $\alpha=1$ and $0<\beta\le 1$, we take $a>\frac{1-\beta}{\beta}$. 
It is not hard to see that an estimate similar to (\ref{E:ist}) is valid with $C$ instead of $\widetilde{C}$.
Now it follows from Corollary \ref{C:gik2} that formula (\ref{E:vost5}) holds.

The proof of Theorem \ref{T:stroussa} is thus completed.

Similar results can be obtained in the case of the left-tail-wing formulas established in \cite{keyBF}. 
We will only forulate the following assertion which is equivalent to Theorem \ref{T:cp}:
\begin{corollary}\label{C:cp1}
Let $C$ be a call pricing function, and let $P$ be the corresponding put pricing function. Then
$$
I(K)=\frac{\log\frac{1}{K}}{\sqrt{T}}\sqrt{\psi\left(\frac{\log P(K)}{\log K}-1\right)}
+O\left(\left(\log\frac{K}{P(K)}\right)^{-\frac{1}{2}}\log\log\frac{K}{P(K)}\right)
$$
as $K\rightarrow 0$, where $\psi(u)=2-4\left(\sqrt{u^2+u}-u\right)$, $u\ge 0$.
\end{corollary}
The equivalence of Theorem \ref{T:cp} and Corollary \ref{C:cp1} can be shown using (\ref{E:fo1}) with 
$a=(\log K)^{-1}\log P(K)-1$.
\section{Stock price distribution densities and pricing functions in stochastic volatility models}\label{S:exam}
This section begins with a trivial example which is the implied volatility in the Black-Scholes model. 
Let us see what can be obtained by applying Theorem \ref{T:stroussa} to the Black-Scholes model, 
and at least make sure that $\displaystyle{\lim_{K\rightarrow\infty}I(K)=\sigma}$. A similar computation was felicitously
called a sanity check in \cite{keyBF}. 

The distribution density of the stock price in the Black-Scholes model is given by
$$
D_T(x)=\frac{\sqrt{x_0e^{rT}}}{\sqrt{2\pi T}\sigma}\exp\left\{-\frac{\sigma^2 t}{8}\right\}
x^{-\frac{3}{2}}\exp\left\{-\left(\log\frac{x}{x_0e^{rT}}\right)^2\left(2T\sigma^2\right)^{-1}\right\}.
$$
This follows from (\ref{E:fof}). Hence, $D_T$ satisfies condition (\ref{E:usl10}) in Theorem \ref{T:stroussa} with 
$$
h(u)=\frac{1}{2T\sigma^2}\left(u-\log\left(x_0e^{rT}\right)\right)^2+\frac{1}{2}u.
$$
It is clear that $h\in SR_2$. Applying Theorem \ref{T:stroussa}, we get
$$
I(K)=\frac{1}{\sqrt{T}}\left[\sqrt{\frac{1}{T\sigma^2}\left(\log\frac{K}{x_0e^{\mu T}}\right)^2+\log K}
-\sqrt{\frac{1}{T\sigma^2}\left(\log\frac{K}{x_0e^{\mu T}}\right)^2-\log K}\right]+O\left(\frac{\log\log K}{\log K}\right)
$$
as $K\rightarrow\infty$. It is not hard to see that right-hand side of the previous asymptotic formula tends to $\sigma$ as $K\rightarrow\infty$.
\\
\\
\it Regularly varying stock price distributions. \rm 
Recall that Pareto-type distributions were defined in Section \ref{S:cBF}. Our next goal is to show that in certain
stock price models with stochastic volatility, the stock price
$X_t$ is Pareto-type distributed for every $t> 0$. 
The models of our interest here are the uncorrelated Stein-Stein, Heston, and Hull-White models. We will first formulate several results 
obtained in \cite{keyGS1,keyGS3,keyGS4}. Recall that the stock price process $X_t$ and the 
volatility process $Y_t$ in the Hull-White  model 
satisfy the following system of stochastic differential equations:
\begin{equation}
\left\{\begin{array}{l}
dX_t=rX_tdt+Y_tX_tdW_t^{*} \\
dY_t=\nu Y_tdt+\xi Y_tdZ_t^{*}.
\end{array}
\right.
\label{E:sde1}
\end{equation}
In (\ref{E:sde1}), $r\ge 0$ is the interest rate, $\nu\in\mathbb{R}^1$, and $\xi> 0$. 
The Hull-White model was introduced in \cite{keyHW}. The volatility process in this model is a geometric Brownian motion.

The Stein-Stein model is defined as follows:
\begin{equation}
\left\{\begin{array}{ll}
dX_t=rX_tdt+\left|Y_t\right|X_tdW_t^{*} \\
dY_t=q\left(m-Y_t\right)dt+\sigma dZ_t^{*}.
\end{array}
\right.
\label{E:SS}
\end{equation}
It was introduced and studied in \cite{keySS}. In this model, 
the absolute value of an Ornstein-Uhlenbeck 
process plays the role of the volatility of the stock. We assume that $r\ge 0$, $q\ge 0$, $m\ge 0$, and $\sigma> 0$. 

The Heston model was developed in \cite{keyH}. It is given by
\begin{equation}
\left\{\begin{array}{ll}
dX_t=rX_tdt+\sqrt{Y_t}X_tdW_t^{*} \\
dY_t=q\left(m-Y_t\right)dt+c\sqrt{Y_t}dZ_t^{*},
\end{array}
\right.
\label{E:H}
\end{equation}
where $r\ge 0$ and $m\ge 0$, and $c\ge 0$. The volatility equation in (\ref{E:H}) is uniquely solvable in the strong sense, 
and the solution $Y_t$ is a positive stochastic process. 
This process is called a Cox-Ingersoll-Ross process. We assume that the processes $W_t^{*}$ and $Z_t^{*}$ in (\ref{E:sde1}), 
(\ref{E:SS}), and (\ref{E:H}) are 
independent Brownian motions under a risk-free probability $\mathbb{P}^{*}$. The initial conditions for the processes 
$X$ and $Y$ are denoted by $x_0$ and $y_0$, respectively.

We will next formulate sharp asymptotic formulas for the 
distribution density $D_t$ of the stock price $X_t$ in special stock price models. These formulas were obtained in 
\cite{keyGS1,keyGS3,keyGS4}.
\begin{enumerate}
\item \it Stein-Stein model. \rm The following result was established in \cite{keyGS4}: There exist positive constants $B_1$,
$B_2$, and $B_3$ such that
\begin{equation}
D_t\left(x_0e^{rt}x\right)=B_1(\log x)^{-\frac{1}{2}}e^{B_2\sqrt{\log x}}x^{-B_3}
\left(1+O\left((\log x)^{-\frac{1}{4}}\right)\right)
\label{E:dop1}
\end{equation}
as $x\rightarrow\infty$. The constants in (\ref{E:dop1}) depend on the model parameters. The constant $B_3$ satisfies $B_3> 2$. 
Explicit formulas for the constants $B_1$,
$B_2$, and $B_3$ can be found in \cite{keyGS4}. It follows from (\ref{E:dop1}), the mean value theorem, and the inequality 
\begin{equation}
e^a-1\le ae^a,\,\,\,0\le a\le 1, 
\label{E:ner}
\end{equation}
that
\begin{equation}
D_t(x)=B_0(\log x)^{-\frac{1}{2}}e^{B_2\sqrt{\log x}}x^{-B_3}
\left(1+O\left((\log x)^{-\frac{1}{4}}\right)\right)
\label{E:dop1s}
\end{equation}
as $x\rightarrow\infty$ where $B_0=\left(x_0e^{rt}\right)^{B_3}$. 
\item \it Heston model. \rm It was shown in \cite{keyGS4} that there exist constants $A_1> 0$, $A_2> 0$, and $A_3> 2$ such that
\begin{equation}
D_t\left(x_0e^{rt}x\right)=A_1(\log x)^{-\frac{3}{4}+\frac{qm}{c^2}}e^{A_2\sqrt{\log x}}x^{-A_3}
\left(1+O\left((\log x)^{-\frac{1}{4}}\right)\right)
\label{E:dopo}
\end{equation}
as $x\rightarrow\infty$. The constants in (\ref{E:dopo}) depend on the model parameters. Explicit expressions for these 
constants can be found in \cite{keyGS4}. 
It follows from (\ref{E:dopo}), the mean value theorem, and (\ref{E:ner}) that
\begin{equation}
D_t(x)=A_0(\log x)^{-\frac{3}{4}+\frac{qm}{c^2}}e^{A_2\sqrt{\log x}}x^{-A_3}
\left(1+O\left((\log x)^{-\frac{1}{4}}\right)\right)
\label{E:dopos}
\end{equation}
as $x\rightarrow\infty$ where $A_0=A_1\left(x_0e^{rt}\right)^{A_3}$.
\item \it Hull-White model. \rm The following asymptotic formula holds for the distribution density of the stock price in the 
Hull-White model (see Theorem 4.1 in \cite{keyGS3}):
\begin{align}
&D_t\left(x_0e^{rt}x\right)=
Cx^{-2}(\log x)^{\frac{c_2-1}{2}}
\left(\log\log x\right)^{c_3} \nonumber \\
&\quad\exp\left\{-\frac{1}{2t\xi^2}\left(\log\left[\frac{1}{y_0}\sqrt{\frac{2\log x}{t}}\right]
+\frac{1}{2}\log\log\left[\frac{1}{y_0}\sqrt{\frac{2\log x}{t}}\right]\right)^2\right\}
\left(1+O\left((\log\log x)^{-\frac{1}{2}}\right)\right)
\label{E:main3}
\end{align}
as $x\rightarrow\infty$. The constants $C$, $c_2$, and $c_3$ have been computed in \cite{keyGS3}. It follows from (\ref{E:main3}),
the mean value theorem, and (\ref{E:ner}) that
\begin{align}
&D_t(x)=
C_0x^{-2}(\log x)^{\frac{c_2-1}{2}}
\left(\log\log x\right)^{c_3} \nonumber \\
&\quad\exp\left\{-\frac{1}{2t\xi^2}\left(\log\left[\frac{1}{y_0}\sqrt{\frac{2\log x}{t}}\right]
+\frac{1}{2}\log\log\left[\frac{1}{y_0}\sqrt{\frac{2\log x}{t}}\right]\right)^2\right\}
\left(1+O\left((\log\log x)^{-\frac{1}{2}}\right)\right)
\label{E:main3s}
\end{align}
as $x\rightarrow\infty$ where $C_0=C\left(x_0e^{rt}\right)^2$.
\end{enumerate}
Equalities (\ref{E:dop1s}), (\ref{E:dopos}), and (\ref{E:main3s}) show that the distribution density $D_t$ of the stock price in the Stein-Stein, Heston, 
and Hull-White model is well-fit by a power law. Indeed, for the Stein-Stein model
we have 
\begin{equation}
D_t(x)\sim x^{-\beta_t}h_t(x),\,\,\,x\rightarrow\infty,
\label{E:su1}
\end{equation}
with 
\begin{equation}
\beta_t=B_3\quad\mbox{and}\quad h_t(x)=B_0(\log x)^{-\frac{1}{2}}e^{B_2\sqrt{\log x}},
\label{E:sus1}
\end{equation}
and it is not hard to see that $h_t\in R_0$. For the Heston model, (\ref{E:su1}) is valid with 
\begin{equation}
\beta_t=A_3\quad\mbox{and}\quad h_t(x)=A_0(\log x)^{-\frac{3}{4}+\frac{qm}{c^2}}e^{A_2\sqrt{\log x}}.
\label{E:sus2}
\end{equation}
The function $h_t$ defined in (\ref{E:sus2}) is a slowly varying function. For the Hull-White model, condition (\ref{E:su1})
holds with $\beta_t=2$ and
\begin{equation}
h_t(x)=C_0(\log x)^{\frac{c_2-1}{2}}
\left(\log\log x\right)^{c_3}\exp\left\{-\frac{1}{2t\xi^2}\left(\log\left[\frac{1}{y_0}\sqrt{\frac{2\log x}{t}}\right]
+\frac{1}{2}\log\log\left[\frac{1}{y_0}\sqrt{\frac{2\log x}{t}}\right]\right)^2\right\}.
\label{E:sus3}
\end{equation}
Here we also have $h_t\in R_0$. Note that the constants $B_3$ and $A_3$ in (\ref{E:sus1}) and (\ref{E:sus2}) depend on $t$ 
(see \cite{keyGS4}).
\begin{remark}\label{R:svrl} \rm
It is also true that the functions $h_t$ in (\ref{E:sus1}) and (\ref{E:sus2}) are slowly varying with remainder \\ 
$g(x)=(\log x)^{-\frac{1}{2}}$. To prove this fact, put
$h_{a,b}(x)=(\log x)^ae^{b\sqrt{\log x}}$ where $b> 0$ and $a\in\mathbb{R}$. Then the function $h_{a,b}$ is slowly varying 
with remainder $g(x)=(\log x)^{-\frac{1}{2}}$. 
This follows from the following asymptotic formula:
\begin{align*}
&\left|\frac{h_{a,b}(\lambda x)}{h_{a,b}(x)}-1\right|
=\frac{(\log x+\log\lambda)^a\left[\exp\left\{b\sqrt{\log x+\log\lambda}-b\sqrt{\log x}\right\}-1\right]
+(\log x+\log\lambda)^a-(\log x)^a}{(\log x)^a} \\
&\quad+O\left((\log x)^{-\frac{1}{2}}\right),\quad x\rightarrow\infty.
\end{align*}
\end{remark}

The next theorem shows that the distribution of the stock price $X_t$ in the Stein-Stein, Heston, and Hull-White models is of Pareto-type.
\begin{theorem}\label{T:cumm}
The following are true for the complementary distribution function $\overline{F}_t$:
\begin{enumerate}
\item Let $t> 0$ and let $\overline{F}_t$ be the complementary distribution function of the stock price $X_t$ in the Stein-Stein model. Then
\begin{equation}
\overline{F}_t(y)\sim y^{-\alpha_t}\tilde{h}_t(y)
\label{E:surv}
\end{equation}
as $y\rightarrow\infty$. In (\ref{E:surv}), $\alpha_t=B_3-1$ and $\tilde{h}_t(y)=\frac{1}{B_3-1}h_t(y)$ where $h_t$ 
is defined in (\ref{E:sus1}).
\item For the Heston model, formula (\ref{E:surv}) holds with $\alpha_t=A_3-1$, $\tilde{h}(y)=\frac{1}{A_3-1}h_t(y)$ 
where $h_t$ is defined in (\ref{E:sus2}).
\item For the Hull-White model, the formula $\overline{F}_t(y)\sim y^{-1}h_t(y)$ holds with $h_t$ defined in (\ref{E:sus3}).
\end{enumerate}
\end{theorem}.

To prove Theorem \ref{T:cumm}, we integrate equalities (\ref{E:dop1s}), (\ref{E:dopos}), and (\ref{E:main3s}) on the interval $[x,\infty)$
and take into account the following theorem due to Karamata: For all $\alpha<-1$ and $l\in R_0$,
$$
\frac{x^{\alpha+1}l(x)}{\int_x^{\infty}t^{\alpha}l(t)dt}\rightarrow-\alpha-1
$$
as $x\rightarrow\infty$ (see Proposition 1.5.10 in \cite{keyBGT}). 

It follows from Theorem \ref{T:cumm} that the Pareto-type index $\alpha_t$ of the stock price $X_t$ in the Stein-Stein model is equal to $B_3-1$. For the Heston model, we have
$\alpha_t=A_3-1$, and for the Hull-Whie model, the Pareto-type index satisfies $\alpha_t=1$.
\section{Asymptotic behavior of call pricing functions in special stochastic volatility models}\label{S:ssvm}
In this section, we obtain sharp asymptotic formulas for call pricing functions in the Hull-White, Stein-Stein, and Heston models. 
The next theorem provides a general asymptotic formula 
for a call pricing function under certain restrictions on the distribution density of the stock price.
\begin{theorem}\label{T:frirego}
Let $C$ be a call pricing function and suppose that the distribution of the stock price $X_T$ admits a density $D_T$. 
Suppose also that 
\begin{equation}
D_T(x)=x^{\beta}h(x)(1+O(\rho(x)))
\label{E:druo}
\end{equation} 
as $x\rightarrow\infty$, where $\beta<-2$, $h$ is a slowly varying function with remainder $g$, and $\rho(x)\downarrow 0$ 
as $x\rightarrow\infty$. Then
\begin{equation}
C(K)=e^{-rT}\frac{1}{(\beta+1)(\beta+2)}K^{\beta+2}h(K)[1+O(\rho(K))+O(g(K))]
\label{E:frirego}
\end{equation}
as $K\rightarrow\infty$.
\end{theorem}

\it Proof. \rm The following assertion (see Theorem 3.1.1 in \cite{keyGS}, or problem 30 on p. 192 in \cite{keyBGT}) 
will be used in the proof:
\begin{theorem}\label{T:abt}
Let $L$ be a slowly varying function with remainder $g$, and let $v$ be a positive function on $(1,\infty)$ such that 
$$
\int_1^{\infty}\lambda^{\epsilon}v(\lambda)d\lambda<\infty\quad\mbox{for some}\quad \epsilon\ge 0.
$$
Then
$$
\int_1^{\infty}v(\lambda)\frac{L(\lambda x)}{L(x)}d\lambda=\int_1^{\infty}v(\lambda)d\lambda+O(g(x))
$$
as $x\rightarrow\infty$.
\end{theorem}

It follows from (\ref{E:druo}) and Theorem \ref{T:abt} that
\begin{align*}
C(K)&=e^{-rT}\int_K^{\infty}(x-K)D_T(x)dx=e^{-rT}\int_K^{\infty}(x-K)x^{\beta}h(x)dx(1+O(\rho(K))) \\
&=e^{-rT}K^{\beta+2}h(K)\int_1^{\infty}(y-1)y^{\beta}\frac{h(Ky)}{h(K)}dy(1+O(\rho(K))) \\
&=e^{-rT}K^{\beta+2}h(K)\int_1^{\infty}(y-1)y^{\beta}dy[1+O(\rho(K))+O(g(K))]
\end{align*}
as $K\rightarrow\infty$. Now it is clear that formula (\ref{E:frirego}) holds, and the proof of Theorem \ref{T:frirego} is thus completed.

Theorem \ref{T:frirego} allows us to characterize the asymptotic behavior of the call pricing function $C(K)$ in the Stein-Stein and 
the Heston model.
\begin{theorem}\label{T:charp}
(a)\,\,\,The following formula holds for the call pricing function $C(K)$ in the Stein-Stein model:
\begin{equation}
C(K)=e^{-rT}\frac{B_0}{\left(1-B_3\right)\left(2-B_3\right)}(\log K)^{-\frac{1}{2}}e^{B_2\sqrt{\log K}}K^{2-B_3}\left(1+O\left((\log K)^{-\frac{1}{4}}\right)\right)
\label{E:now0}
\end{equation}
as $K\rightarrow\infty$. The constants in (\ref{E:now0}) are the same as in (\ref{E:dop1s}). \\
(b)\,\,\,The following formula holds for for the call pricing function $C(K)$ in the Heston model:
\begin{equation}
C(K)=e^{-rT}\frac{A_0}{\left(1-A_3\right)\left(2-A_3\right)}(\log K)^{-\frac{3}{4}
+\frac{qm}{c^2}}e^{A_2\sqrt{\log K}}K^{2-A_3}\left(1+O\left((\log K)^{-\frac{1}{4}}\right)\right)
\label{E:now00}
\end{equation}
as $K\rightarrow\infty$. The constants in (\ref{E:now00}) are the same as in (\ref{E:dopos}).
\end{theorem}

It is not hard to see that Theorem \ref{T:charp} follows from (\ref{E:dop1s}), (\ref{E:dopos}), Remark \ref{R:svrl}, and Theorem \ref{T:frirego}.

Next we turn our attention to the Hull-White model. Note that Theorem \ref{T:charp} can not be applied in this case
since for the Hull-White model we have $\beta=-2$. Instead, we will employ an asymptotic formula for fractional 
integrals established in \cite{keyGS3}
(see Theorem 3.7 in \cite{keyGS3}). A special case of this formula is as follows: Let $b(x)=B(\log x)$ be a 
positive increasing function on $[c,\infty)$ 
with $B^{\prime\prime}(x)\approx 1$ as $x\rightarrow \infty$. Then
\begin{equation}
\int_K^{\infty}\exp\{-b(x)\}dx
=\frac{\exp\{-b(K)\}}{b^{\prime}(K)}\left(1+O\left(\left(\log K\right)^{-1}\right)\right)
\label{E:formb}
\end{equation}
as $K\rightarrow\infty$. Formula (\ref{E:formb}) will be used in the proof of the following result:
\begin{theorem}\label{T:rhw}
Let $C$ be a call pricing function, and suppose that the distribution of the stock price $X_T$ admits a density $D_T$. Suppose also that
\begin{equation}
D_T(x)=x^{-2}\exp\{-b(\log x)\}(1+O(\rho(x)))
\label{E:druo}
\end{equation} 
as $x\rightarrow\infty$. Here the function $b$ is positive, increasing on $[c,\infty)$ for some $c> 0$, and such that
the condition $b(x)=B(\log x)$. 
Moreover, $B^{\prime\prime}(x)\approx 1$ as $x\rightarrow \infty$, and $\rho(x)\downarrow 0$ as $x\rightarrow\infty$. Then
\begin{equation}
C(K)=e^{-rT}\frac{\exp\{-b(\log K)\}\log K}{B^{\prime}(\log\log K)}\left[1+O\left(\left(\log\log K\right)^{-1}\right)+O(\rho(K))\right]
\label{E:frair}
\end{equation}
as $K\rightarrow\infty$.
\end{theorem}

\it Proof. \rm We have
\begin{align*}
&C(K)=e^{-rT}\left[\int_K^{\infty}xD_T(x)dx-K\int_K^{\infty}D_T(x)dx\right] \\
&=e^{-rT}\left[\int_K^{\infty}x^{-1}\exp\{-b(\log x)\}dx
-K\int_K^{\infty}x^{-2}\exp\{-b(\log x)\}dx\right](1+O(\rho(K)) \\
&=e^{-rT}\int_K^{\infty}x^{-1}\exp\{-b(\log x)\}dx(1+O(\rho(K))+O\left(\exp\{-b(\log K)\}\right) \\
&=e^{-rT}\int_{\log K}^{\infty}\exp\{-b(u)\}dx(1+O(\rho(K))+O\left(\exp\{-b(\log K)\}\right).
\end{align*}
Using (\ref{E:formb}) we get
\begin{align}
&C(K)=e^{-rT}\frac{\exp\{-b(\log K)\}}{b^{\prime}(\log K)}\left(1+O\left(\left(\log\log K\right)^{-1}\right)\right)
[1+O(\rho(K)]+O\left(\exp\{-b(\log K)\}\right) \nonumber \\
&=e^{-rT}\frac{\exp\{-b(\log K)\}\log K}{B^{\prime}(\log\log K)}\left[1+O\left(\left(\log\log K\right)^{-1}\right)
+O(\rho(K))\right]+O\left(\exp\{-b(\log K)\}\right). 
\label{E:fra} 
\end{align}
Since $B^{\prime}(x)\approx x$ as $x\rightarrow\infty$, (\ref{E:fra}) implies (\ref{E:frair}).

The next statement characterizes the asymptotic behavior of a call pricing function in the Hull-White model.
\begin{theorem}
Let $C$ be a call pricing function in the Hull-White model. Then 
\begin{align}
&C(K)=4T\xi^2C_0e^{-rT}(\log K)^{\frac{c_2+1}{2}}\left(\log\log x\right)^{c_3-1} \nonumber \\
&\quad\exp\left\{-\frac{1}{2T\xi^2}\left(\log\left[\frac{1}{y_0}\sqrt{\frac{2\log K}{T}}\right]
+\frac{1}{2}\log\log\left[\frac{1}{y_0}\sqrt{\frac{2\log K}{T}}\right]\right)^2\right\}                      
\left(1+O\left((\log\log K)^{-\frac{1}{2}}\right)\right)
\label{E:now1}
\end{align}
as $K\rightarrow\infty$. The constants in (\ref{E:now1}) are the same as in formula (\ref{E:dop1s}).
\end{theorem}

\it Proof. \rm We will employ Theorem \ref{T:rhw} in the proof. It is not hard to see using (\ref{E:main3s}) 
that formula (\ref{E:druo}) holds for the distribution density $D_T$
of the stock price in the Hull-White model. Here we choose the functions $b$, $B$, and $\rho$ as follows:
$$
b(u)=-\log C_0-\frac{c_2-1}{2}\log u-c_3\log\log u+\frac{1}{2T\xi^2}\left(\log\left[\frac{1}{y_0}\sqrt{\frac{2u}{T}}\right]
+\frac{1}{2}\log\log\left[\frac{1}{y_0}\sqrt{\frac{2u}{T}}\right]\right)^2,
$$
$$
B(u)=-\log C_0-\frac{c_2-1}{2}u-c_3\log u+\frac{1}{2T\xi^2}\left[\log\frac{1}{y_0}\sqrt{\frac{2}{T}}+\frac{1}{2}u
+\frac{1}{2}\log\left(\log\frac{1}{y_0}\sqrt{\frac{2}{T}}+\frac{1}{2}u\right)\right]^2,
$$
and $\rho(x)=(\log\log x)^{-\frac{1}{2}}$. It is clear that $B^{\prime\prime}(u)\approx 1$ and $B^{\prime}(u)\approx u$ as $u\rightarrow\infty$.
Moreover, using the mean value theorem, we obtain the following estimate:
\begin{equation}
\frac{1}{B^{\prime}(\log\log K)}-\frac{4T\xi^2}{\log\log K}=O\left((\log\log K)^{-2}\right)
\label{E:now2}
\end{equation}
as $K\rightarrow\infty$. Next, taking into account (\ref{E:frair}) and (\ref{E:now2}), we see that (\ref{E:now1}) holds.
\begin{remark}\label{R:final} \rm Theorem \ref{T:generu} and formulas (\ref{E:now0}), (\ref{E:now00}), and (\ref{E:now1}) 
can be used to obtain the asymptotic formulas with error estimates for the implied volatility in the Hull-White, Stein-Stein, and Heston models 
established in \cite{keyGS2} and \cite{keyGS4}.
\end{remark}
\section{Appendix}\label{A:a} 
We will next prove the characterization of call pricing functions that was formulated in the introduction. 
The following well-known fact from the theory of convex functions will be used in the proof: 
If $U(x)$ is a convex function on 
$(0,\infty)$, then the second (distributional) derivative $\mu$ of the function $U$ is a locally finite Borel measure on $(0,\infty)$ and any 
such measure is the second derivative of a convex function $U$ which is unique up to the addition of an affine function.

Let $C$ be a pricing function and denote by $\mu_T$ the distribution of the random variable $X_T$. Then the second distributional derivative of the function 
$K\mapsto e^{rT}C(T,K)$ coincides with the measure $\mu_T$. Our goal is to establish that conditions 1-5 in the
characterization of call pricing functions hold. It is clear that conditions 1 and 4 follow from the definitions. Condition 2 can be established 
using the equivalence of (\ref{E:mapr}) and (\ref{E:vo}).
We will next prove that Condition 3 holds. This condition is equivalent to the following inequality:
\begin{equation}
\int_0^{\infty}(x-K)^{+}d\mu_S\left(e^{rS}x\right)\le\int_0^{\infty}(x-K)^{+}d\mu_T\left(e^{rT}x\right)
\label{E:kelre}
\end{equation}
for all $K\ge 0$ and $0\le S\le T<\infty$. The previous inequality can be established by taking into account 
the fact that the process $e^{-rt}X_t$, $t\ge 0$, is a martingale and applying Jensen's inequality.
Finally, Condition 5 follows from the estimate
$$
C(T,K)\le e^{-rT}\int_K^{\infty}xd\mu_T(x)
$$ 
and (\ref{E:vo}). This proves the necessity in the characterization of call pricing functions.

To prove the sufficiency, let us assume that $C$ is a function such that Conditions 1-5 hold. Consider the family of Borel probability measures 
$\left\{\nu_T\right\}_{T\ge 0}$ on
$\mathbb{R}$ defined as follows: For every $T\ge 0$, $\nu_T(A)=\mu_T\left(e^{rT}A\right)$ if $A$ is a Borel subset of $[0,\infty)$, and $\nu_T((-\infty,0))=0$. 
In this definition, the symbol $\mu_T$ 
stands for the second distributional derivative of the function
$K\mapsto e^{rT}C(T,K)$. Since Condition 3 is equivalent to (\ref{E:kelre}), we have
\begin{equation}
\int_{\mathbb{R}}(x-K)^{+}d\nu_S(x)\le\int_{\mathbb{R}}(x-K)^{+}d\nu_T(x)
\label{E:convo1}
\end{equation}
for all $K\ge 0$ and $0\le S\le T<\infty$. Let $\varphi$ be a non-decreasing convex function on $[0,\infty)$ and denote by $\eta$ its second distributional derivative. 
Then we have
$\displaystyle{\varphi(x)=\int_0^{\infty}(x-u)^{+}d\eta(u)}+ax+b$ for all $x\ge 0$, where $a$ and $b$ are some constants. Next using (\ref{E:convo1}) and (\ref{E:vo}), we obtain
\begin{equation}
\int_{\mathbb{R}}\varphi(x)d\nu_S(x)\le\int_{\mathbb{R}}\varphi(x)d\nu_T(x)
\label{E:convo2}
\end{equation}
for all $K\ge 0$, $0\le S\le T<\infty$. Hence, the family $\left\{\nu_T\right\}_{T\ge 0}$ is increasing in the convex ordering. 
The reasoning leading from (\ref{E:convo1}) to (\ref{E:convo2}) is known
(see \cite{keyFS}, or Appendix 1 in \cite{keyCM1}). 
By Kellerer's theorem (Theorem 3 in \cite{keyK}), (\ref{E:convo2}) and (\ref{E:vo}) imply the existence of a filtered 
probability space $\left(\Omega,{\cal F},{\cal F}_t,\mathbb{P}^{*}\right)$ and of a Markov $\left({\cal F}_t,\mathbb{P}^{*}\right)$-martingale $Y$ 
such that the distribution of $Y_T$ coincides with the measure $\nu_T$ for every $T\ge 0$. Now put $X_T=e^{rT}Y_T$, $T\ge 0$. 
It follows that the measure $\mu_T$ is the distribution of the random variable $X_T$ 
for every $T\ge 0$. This produces a stock price process $X$ such that the process $e^{-rt}X_t$ is a martingale. Next, we see that Condition 4
implies $\mu_0=\delta_{x_0}$, and hence $X_0=x_0$ $\mathbb{P}^{*}$-a.s. Taking into account inequality (\ref{E:vo}), 
we define the following function:
\begin{equation}
V(T,K)=\int_K^{\infty}xd\mu_T(x)-K\int_K^{\infty}d\mu_T(x)=\mathbb{E}^{*}\left[\left(X_T-K\right)^{+}\right].
\label{E:c3}
\end{equation}
It is clear that the second distributional derivative of the function $K\mapsto V(T,K)$
coincides with the measure $\mu_T$. Therefore, 
$e^{rT}C(T,K)=V(T,K)+a(T)K+b(T)$ for all $T\ge 0$ and $K\ge 0$, where the functions $a$ and $b$ do not depend on $K$. Since 
$\displaystyle{\lim_{K\rightarrow\infty}C(T,K)=0}$ (Condition 5) and $\displaystyle{\lim_{K\rightarrow\infty}C(T,K)=0}$, we see that 
$a(T)=b(T)=0$, and hence $C(T,K)=e^{-rT}V(T,K)$. It follows from (\ref{E:c3}) that $C$ is a call pricing function.

\end{document}